\documentclass[conference]{IEEEtran}
\IEEEoverridecommandlockouts
\usepackage{cite}
\usepackage{amsmath,amssymb,amsfonts}
\usepackage{algorithmic}
\usepackage{graphicx}
\usepackage{textcomp}
\usepackage{xcolor}
\def\BibTeX{{\rm B\kern-.05em{\sc i\kern-.025em b}\kern-.08em
    T\kern-.1667em\lower.7ex\hbox{E}\kern-.125emX}}

\begin{document}

\title{Experimental Analysis of Harvested Energy and Throughput Trade-off in a Realistic SWIPT System\\
\thanks{This work has been partially supported by the EPSRC of UK, under grants EP/P003885/1 and EP/R511547/1}
}

\author{\IEEEauthorblockN{Junghoon Kim}
\IEEEauthorblockA{\textit{EEE Department} \\
\textit{Imperial College London}\\
London, United Kingdom \\
junghoon.kim15@imperial.ac.uk}
\and
\IEEEauthorblockN{Bruno Clerckx}
\IEEEauthorblockA{\textit{EEE Department} \\
\textit{Imperial College London}\\
London, United Kingdom \\
b.clerckx@imperial.ac.uk}

\and
\IEEEauthorblockN{Paul D. Mitcheson}
\IEEEauthorblockA{\textit{EEE Department} \\
\textit{Imperial College London}\\
London, United Kingdom \\
paul.mitcheson@imperial.ac.uk}
}

\maketitle

\begin{abstract}
We build a realistic Simultaneous Wireless Information and Power Transfer (SWIPT) prototype and experimentally analyse the harvested energy and throughput trade-off. Both time-switching and power splitting receiver architectures are implemented, and the performance comparison is carried out. Systematic SWIPT transmission signal design methods are also considered and implemented on the prototype. The harvested energy-throughput (E-T) performance with different transmission signal designs, modulation schemes, and receiver architectures are evaluated and compared. The combination of the power splitting receiver architecture and the \textit{superposition} transmission signal design technique shows significant expansion of the E-T region. The experimental results fully validate the observations predicted from the theoretical signal designs and confirm the benefits of systematic signal designs on the system performance.  The observations give important insights on how to design a practical SWIPT system. 
\end{abstract}


\section{Introduction}
Simultaneous Wireless Information and Power Transfer (SWIPT) is becoming an emerging research area that makes use of radio waves for the joint purpose of Wireless Information Transfer (WIT) and Wireless Power Transfer (WPT). Due to the growing interest in low-power wireless devices, such as Internet-of-Things (IoT) devices, SWIPT is attracting attention as a technology that can energize and communicate with those devices.

\par
There is a fundamental tradeoff between information transfer rates and energy delivery efficiency, so-called rate-energy (R-E) tradeoff, because of the different objectives of WPT and WIT. Many prior studies on SWIPT have mainly focused on characterizing this R-E tradeoffs in various environments, such as MIMO broadcasting channels \cite{Zhang2013}, interference channels \cite{Park2013}, and broadband systems \cite{Huang2013}.

The system architectures of SWIPT transmitters and receivers also have a remarkable impact on the R-E tradeoff and system performance. Receiver architecture with time-switching (TS) and power-splitting (PS) schemes, in which information decoders and energy harvesters are separated or integrated, have been proposed,  and the R-E tradeoff of each option has been analyzed in \cite{Zhou2013}.  Both TS and PS schemes are common architectures that are widely used in the SWIPT literature \cite{Clerckx2019}.

Another line of studies on SWIPT examines techniques that extend the R-E region under given wireless channel conditions. To this end, the vast majority of research efforts have been devoted to the design of efficient energy harvesters (rectennas). In addition, analytical models for the nonlinear behavior of the rectenna have been studied \cite{Clerckx2019}, and efficient transmission signal design techniques have been proposed based on these models \cite{Clerckx2018d}.

\par
WPT is an important building block of the SWIPT system, and expanding the R-E region of the SWIPT architecture relies heavily on the efficiency of the WPT system's design. Recently, significant WPT performance improvements using systematic channel-adaptive waveform design and optimization have been confirmed through the analytical models, simulations and experiments \cite{Clerckx2016}, \cite{Kim2018}. The systematic waveform design method for WPT of \cite{Clerckx2016} has been successfully extended to SWIPT signal design and has been shown to significantly expand the R-E region \cite{Clerckx2018d}.

\begin{figure*}[!h]
	\centering
	\includegraphics[width=0.75\textwidth]{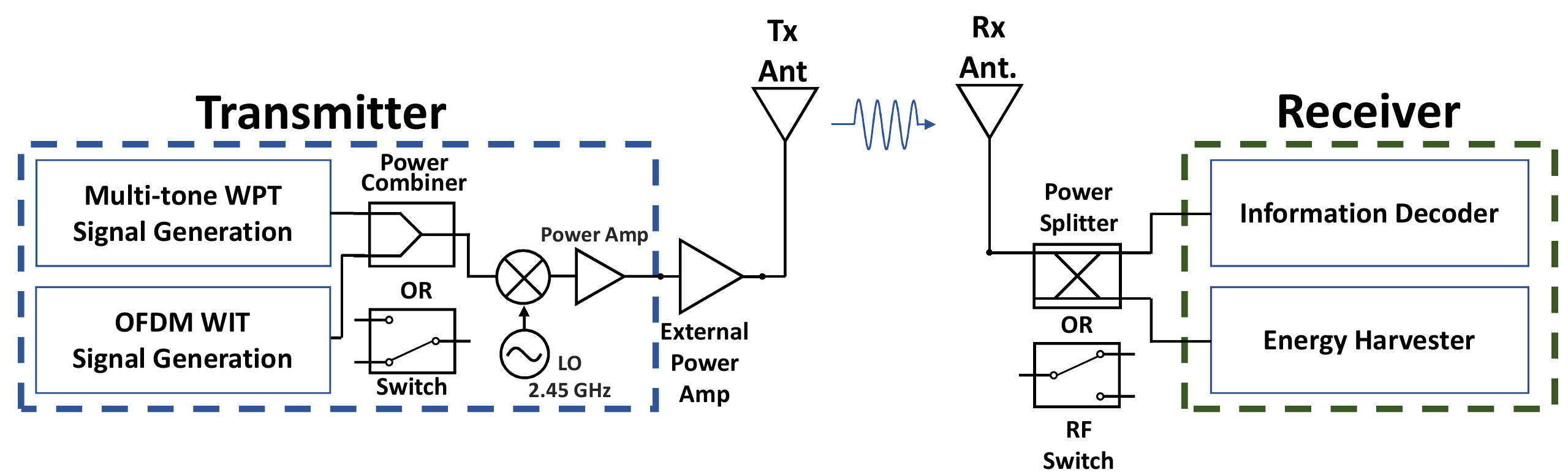}
	\caption{SWIPT prototype system diagram.}
	\label{system_diagram}
\end{figure*}

\par
In this paper, we first establish a SWIPT testbed system as a realistic prototype using software defined radio (SDR) equipment. The system includes TS and PS receiver architectures with information decoder and energy harvester, and it also incorporates a feature of systematic transmission signal design inspired by \cite{Clerckx2018d}. The systematic transmission signal design feature contains various functions, such as the generation of a WIT signal based on 802.11g WiFi standard, generation of an 8-tone multisine WPT signal, and combination of both WIT and WPT signals with time-sharing and power-combining manner. Various modulation schemes are additionally considered for performance evaluation of the R-E region according to the modulation scheme. Then, we experimentally analyze the tradeoff between harvested energy and throughput with various transmission signal design techniques. We seek to evaluate the benefits that can be gained from the systematic signal design technique. We also compare the measurement results with existing theoretical results in \cite{Clerckx2018d} to confirm that the analytical models are in line with experimental measurements.

\par
The rest of this paper is organized as follows. In section \ref{system}, we provide a brief introduction to our system architecture and signal design methods. Experimental performance evaluation results and analysis is presented in section \ref{analysis}. Then, section \ref{conclusion} summaries the work and discusses the future plan. 

\section{Signals and System Design} \label{system}
In this section, we introduce the operation schemes of the different building blocks in our SWIPT prototype including receiver architectures and transmission signal designs used in our experiments. We then provide more details about the implementation of the SWIPT prototype. Fig.\ref{system_diagram} shows an overall system diagram of the SWIPT prototype system.

\subsection{Receiver Architecture}
Various receiver architectures in which information decoder (ID) and energy harvester (EH) are integrated have been proposed in prior researches \cite{Clerckx2019}. We consider an implementation-friendly ID and EH architecture based on \textit{time-switching} (TS) and \textit{power-splitting} (PS). The right side of Fig.\ref{system_diagram} shows the receiver structure. 

\par
The \textit{time-switching (TS) receiver} is equipped with an RF switch behind the receiving antenna, which transfers the received signal to the energy harvester or the information decoder in a time division multiplexing manner. The Rx time-switching ratio, $\alpha_{rx}$, varies from 0 to 1, which refers to the fraction of the time the receiving antenna and energy harvester are connected through the RF switch within one second time slot. During the remaining time $1-\alpha_{rx}$, the receiving antenna is connected to the information decoder. The \textit{power splitting (PS) receiver} splits the received signals into two streams, where one with a power splitting ratio $\rho_{rx}$ is transferred to the energy harvester and the other with power splitting ratio $1-\rho_{rx}$ is transferred to the information decoder. 

\subsection{Transmission Signal Design}\label{signal_design}
\par
The systematic SWIPT transmission signal can be designed through a combination of the information (WIT) and the power (WPT) signals \cite{Clerckx2018d}. We implement two different types of systematic signal design methods. One is called the \textit{time-sharing} method that transmits WPT and WIT signals in a time-division multiplexing manner. The other is called the \textit{superposition} method that superimposes both WPT and WIT signals with a certain power ratio. The transmission SWIPT signals of both methods $x_{ts}$ and $x_{sp}$ at time $t$ are written as

\begin{eqnarray}
 x_{ts}(t) &=& \left\{\begin{matrix} x_{P}(t) & 0 \le t \le \alpha_{tx} \\ x_{I}(t) & \alpha_{tx} < t \le 1 . \end{matrix}\right. \\
 x_{sp}(t) &=& \sqrt{\rho_{tx}} x_{P}(t) + \sqrt{1-\rho_{tx}} x_{I}(t).
\end{eqnarray}
where $x_{P}(t)$ and $x_{I}(t)$ represent WPT and WIT signal respectively, and $\alpha_{tx}$ is a tx time-sharing ratio with $0 \le \alpha_{tx} \le 1$ and $\rho_{tx}$ is tx power combining ratio with $0 \le \rho_{tx} \le 1$. 
\par
\begin{figure}[!h]
	\centering
	\includegraphics[width=0.5\textwidth]{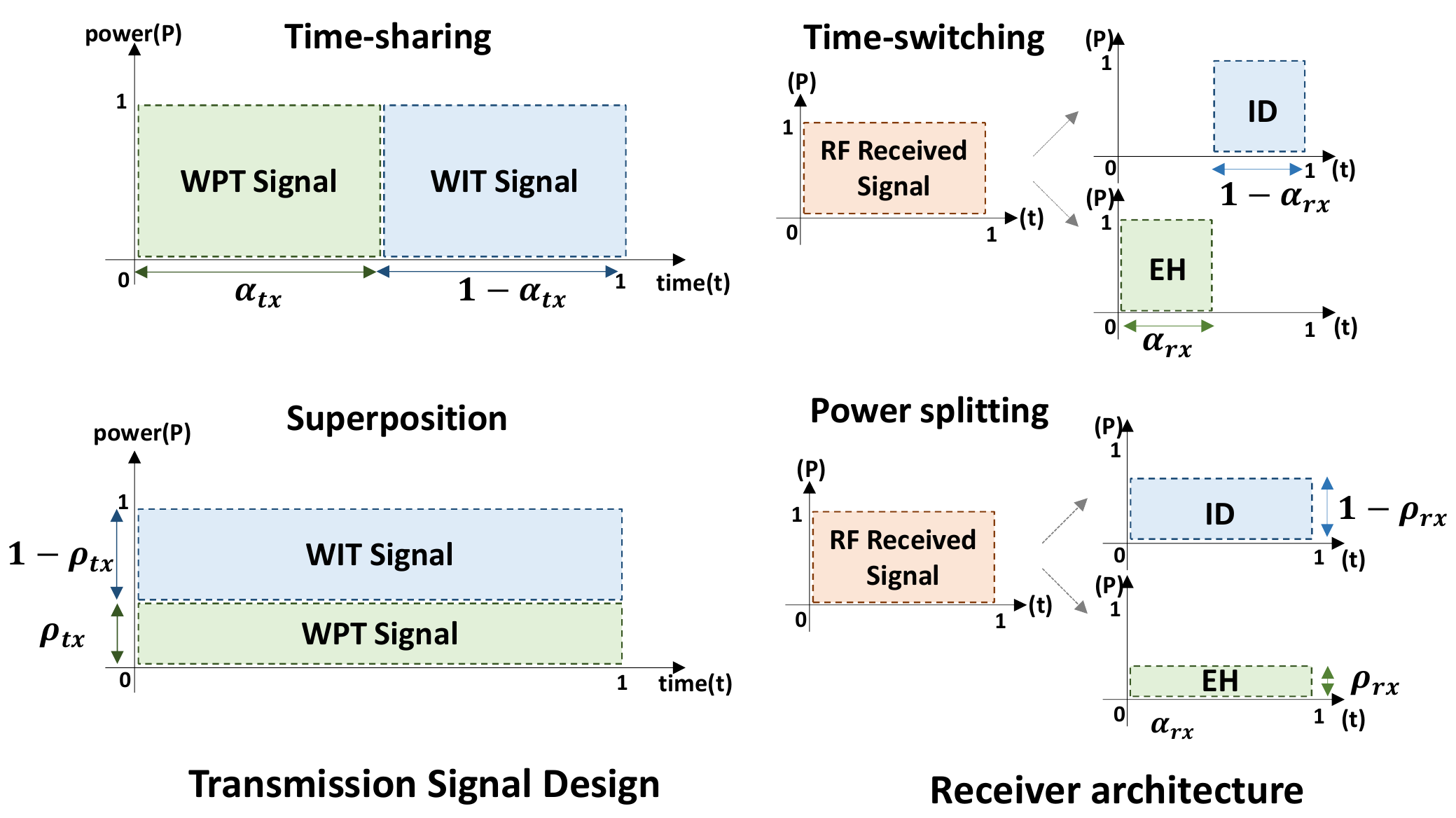}
	\caption{Signal structures at the transmitter and receiver}
	\label{txrxsignal}
\end{figure}

Fig.\ref{txrxsignal} shows the signal structures of the different transmission signal designs and receiver structures and also indicates the roles of parameters $\alpha$ and $\rho$.
\par
\begin{figure}[!h]
	\centering
	\includegraphics[width=0.5\textwidth]{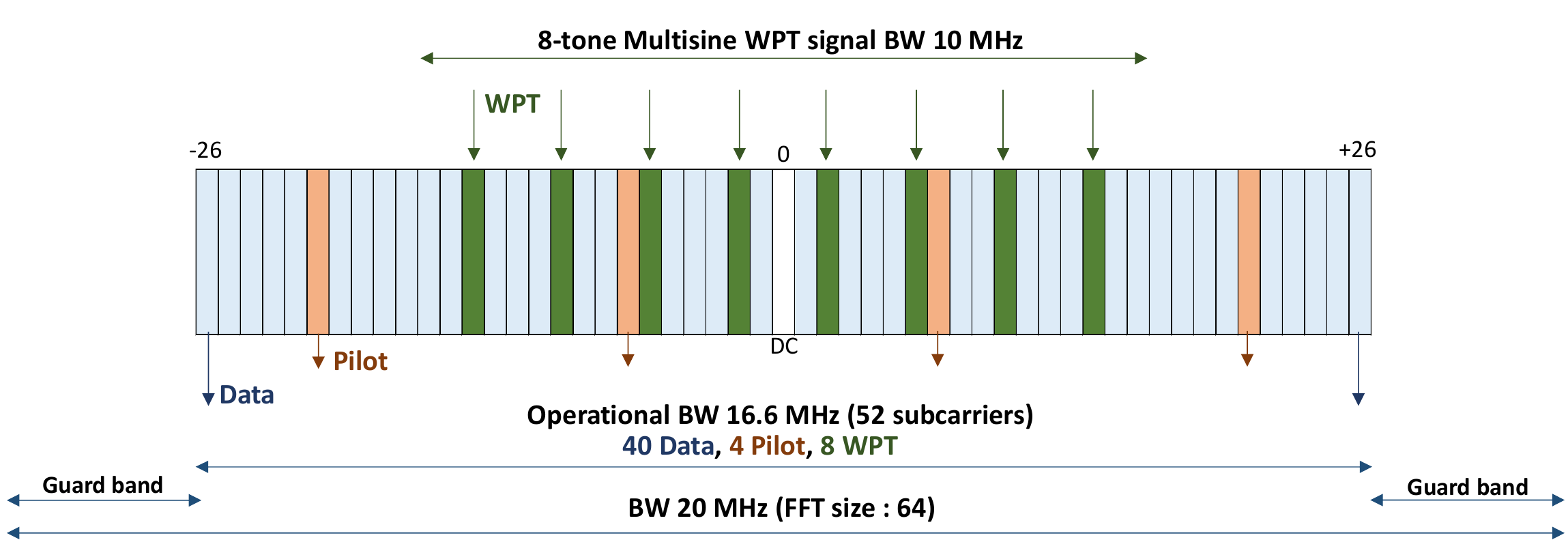}
	\caption{Subcarriers of WPT and WIT combined \textit{superposition} SWIPT signal (based on 802.11g).}
	\label{subcarrier}
\end{figure}

We choose to use an evenly spaced and uniformly power-allocated 8-tone multi-sine waveform with 10 MHz bandwidth as a WPT signal. This waveform is chosen because it has been verified to significantly improve the overall end-to-end efficiency over single-tone waveforms in \cite{Kim2018}. We also choose to use a slightly modified OFDM signal based on WiFi 802.11g standard as WIT signal. This OFDM signal is different from the existing 802.11g standard signal in such a way that it does not use eight subcarriers out of 64 subcarriers and reserve those eight subcarriers for the WPT waveform. This differs a bit from the original superposition approach used in \cite{Clerckx2018d}, where WPT and WIT waveforms are superimposed on the same subcarriers. The approach in \cite{Clerckx2018d} nevertheless requires the WPT signal to be cancelled at the information receiver. However, in this preliminary SWIPT experiment, we have decided not to use overlapping subcarriers for WIT and WPT in order to reduce the implementation complexity (i.e. it is challenging to demodulate if the WPT and WIT signals are overlapped). Therefore, the remaining eight subcarriers do not carry information and are used to transmit the WPT 8-tone multi-sine waveform. \textit{Superposition} signal generation and demodulation techniques, including WPT signal cancellation, remain as future work. Fig.\ref{subcarrier} shows the subcarrier mapping of the WPT and WIT signals of the \textit{superposition} signal.


\par
Because eight subcarriers are not used in the WIT signal, the maximum data rate is also slightly reduced. The maximum data rate according to modulation type with 3/4 coding rate is as follows: 7.5 Mbps for BPSK, 15 Mbps for QPSK, 30Mbps for 16 QAM, and 45Mbps for 64QAM.

\subsection{SWIPT Prototype Implementation}
\par
We implemented the Single-Input Single-Output (SISO) point-to-point SWIPT system using Software Defined Radio (SDR) prototyping equipment. The prototype system operates at 2.4GHz carrier frequency, which is the same as WiFi in the ISM band. A set of National Instrument (NI) FlexRIO (PXI-7966R) and transceiver module (NI 5791R), and a host controller (PXIe-8133) is used as a SWIPT transmitter. The transmitter includes features to generate the WPT and WIT waveforms and to combine them into SWIPT transmission signals in a time-sharing or superposition manner.
\par
On the receiver side, a power splitter or an RF switch is located directly behind the receiving antenna to distribute the received signals to both information decoder (ID) and energy harvester (EH) blocks in power splitting (PS) and time-switching (TS) manner, respectively.  The EH block is a simple single diode rectifier to convert the received RF signals to DC power. The ID block is composed of a separate hardware having the same configuration as the transmitter to demodulate information from the received signal.
\par
In both transmitter and receiver, the power-combining ($\rho_{tx}$), and time-sharing ($\alpha_{tx}$),  and power-splitting ($\rho_{rx}$), and time-switching ($\alpha_{rx}$) ratios are designed to vary from 0 to 1 with 0.1 step. The maximum transmit power is set to 35dBm and the receiver sensitivity of the information decoder (ID) is below -80dBm, which is set to be similar to commercial low power WiFi equipment.

\section{Experimental Performance Analysis}\label{analysis}
\par 
We experimentally evaluate the harvested energy-throughput (E-T) region for various signal design techniques. The E-T region represents the relationship between harvested energy and data throughput measured through the experiment, as opposed to the R-E region shown in prior studies which describe the theoretically achievable harvested energy and the data rate. The performance of the systematic signal design techniques including the \textit{superposition} and the \textit{time-sharing} is measured, and compared with the performance when using the WIT-only signal. In addition, the performance of various modulations for the OFDM waveform is also evaluated. 
\par
The experiment was carried out by setting the received RF power to be fixed at around -20dBm at the receiving antenna. The reason why we choose -20dBm receiving power is that firstly the \textit{systematically designed} waveforms are a technique that improves WPT performance by using nonlinearity of diode in the low power region below 0dBm and secondly the received power below -20dBm is not meaningful because the RF-to-DC conversion efficiency of a rectifier is too low.

\subsection{Performance of Systematically Designed Waveforms}
\par
We first carried out the experiment with a WIT-only signal for both PS and TS receiver as a baseline. The WIT signal is 802.11g standard based OFDM signal as mentioned in section \ref{system} and QPSK modulation is used. Then, we also carried out the experiment with  \textit{systematically designed} SWIPT signals which consist of 8-tone multi-sine WPT signal and QPSK modulated OFDM WIT signal but combined by different methods as \textit{superposition} and \textit{time-sharing} respectively. The \textit{superposition} transmission signal was experimented with a PS receiver, and the \textit{time-sharing} transmission signal was experimented with a TS receiver because the opposite combination of transmission signals and receiver architectures would adversely affect both the harvested energy and throughput performance.
\par
\begin{figure}[!h]
	\centering
	\includegraphics[width=0.42\textwidth]{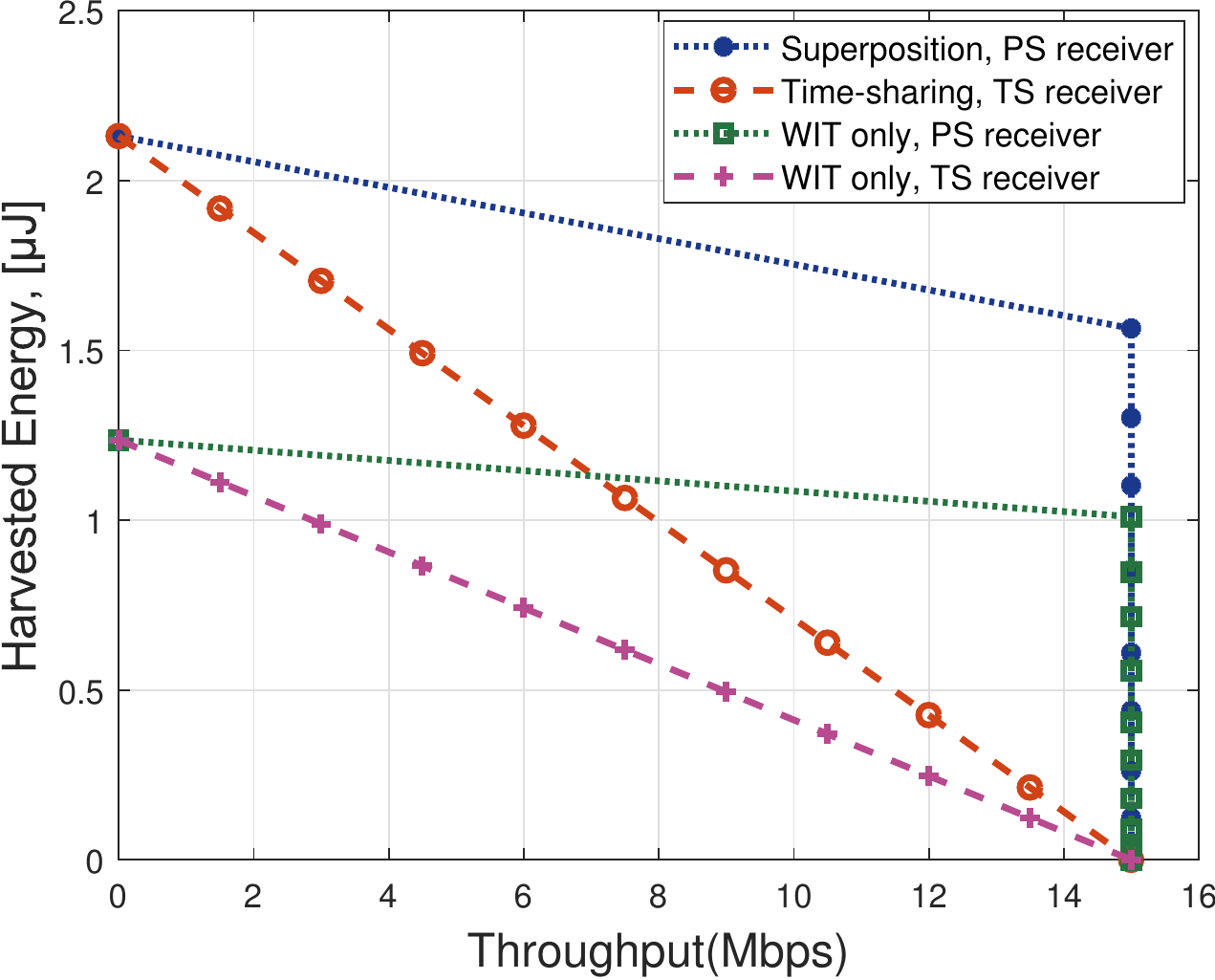}
	\caption{Harvested Energy-Throughput tradeoff with designed signals and WIT only signals in both PS and TS receiver.}
	\label{spvsofdm}
\end{figure}

We measure the output voltage of the EH and bit-error rate (BER) at the ID. Throughput is calculated as $(1-BER) \times max rate$. The experiment ran for one second each and repeated 300 times, the averaged results of energy-throughput tradeoff is shown in Fig.\ref{spvsofdm}.

\par
First, the results show that the energy-throughput (E-T) regions of PS receiver test cases are much larger than TS receiver test cases for both \textit{superposition} and WIT only SWIPT signals. Moreover, the shape of the E-T region almost looks like that of an ideal receiver (that assumes to be able to decode information and harvest energy from the same signal) \cite{Clerckx2019}. Those results are explained by the fact that the SNR at the information decoder is large enough to decode the received signal even in the lowest received power case. Since the RF input power to the receiving antenna is fixed at -20dBm for the operation of the energy harvester, the minimum power of the WIT signal input to the ID of PS receiver is -40 dBm in the worst case ( $\rho_{tx} = 0.9, \rho_{rx} = 0.9$), except for the case of $\rho_{tx} = 1$ or $\rho_{rx} = 1$. An input power of -40dBm to the ID is strong enough to decode the information for the given receiver specification (receiver sensitivity of ID is less than -80dBm). In other words, the ID can successfully receive information with a very low portion of the input power, but in the case of the TS receiver architecture, all input power is sent to the ID for a certain period of $\alpha_{rx}$ and the EH cannot harvest energy during that period. This is why the E-T region of TS receiver test cases are much smaller than the test cases of the PS receiver. Those results advocate that the PS receiver architecture outperforms the TS architecture in the SWIPT system which equipped ID with high-receiver sensitivity such as conventional low-power Wi-Fi or RF information receivers. This observation is consistent with the theoretical analysis of the R-E performance of \textit{superposition} signal with PS architecture in various Signal-to-Noise ratio (SNR) environment. The theoretical analysis has shown that the combination of PS architecture and \textit{superposition} signal outperforms the TS and \textit{time-sharing} combination only in the high SNR ($>$40dB) environment, and the results are reversed in the low SNR ($<$30dB) environment \cite{Clerckx2018d}. Since the ID has a good receiver sensitivity, the experiment is effectively conducted in the high SNR regime, and we can confirm and validate experimentally the high SNR behavior predicted from theory in \cite{Clerckx2018d}. Performing experiments at low SNR using a different type of ID remains as a future task. 

 

\par
Second, \textit{systematically designed} SWIPT signals show better energy transfer performance for both PS and TS receiver architecture. By using an 8-tone multisine signal as a WPT signal, the maximum achievable harvested energy is increased by 72\% compared to single tone WPT signal, thereby the E-T region is significantly expanded. This key observation comes from the nonlinear behaviour of the EH receiver \cite{Clerckx2016}. In SWIPT, as well as in WPT, we therefore confirm experimentally that a systematic signal design plays a crucial role to improve the harvested energy performance of the system. Furthermore, the experimental results have shown that the E-T region can be adjusted through the appropriate combination of WIT and WPT signals. This observation is inline with the observations made from the WPT literature which has shown the beneficial role of multisine WPT signal \cite{Kim2018}, and is also inline with the theoretical analysis of systematic signal design techniques \cite{Clerckx2018d}\cite{Clerckx2016}.

\subsection{Comparison of different modulations}
\par
\begin{figure}[!h]
	\centering
	\includegraphics[width=0.42\textwidth]{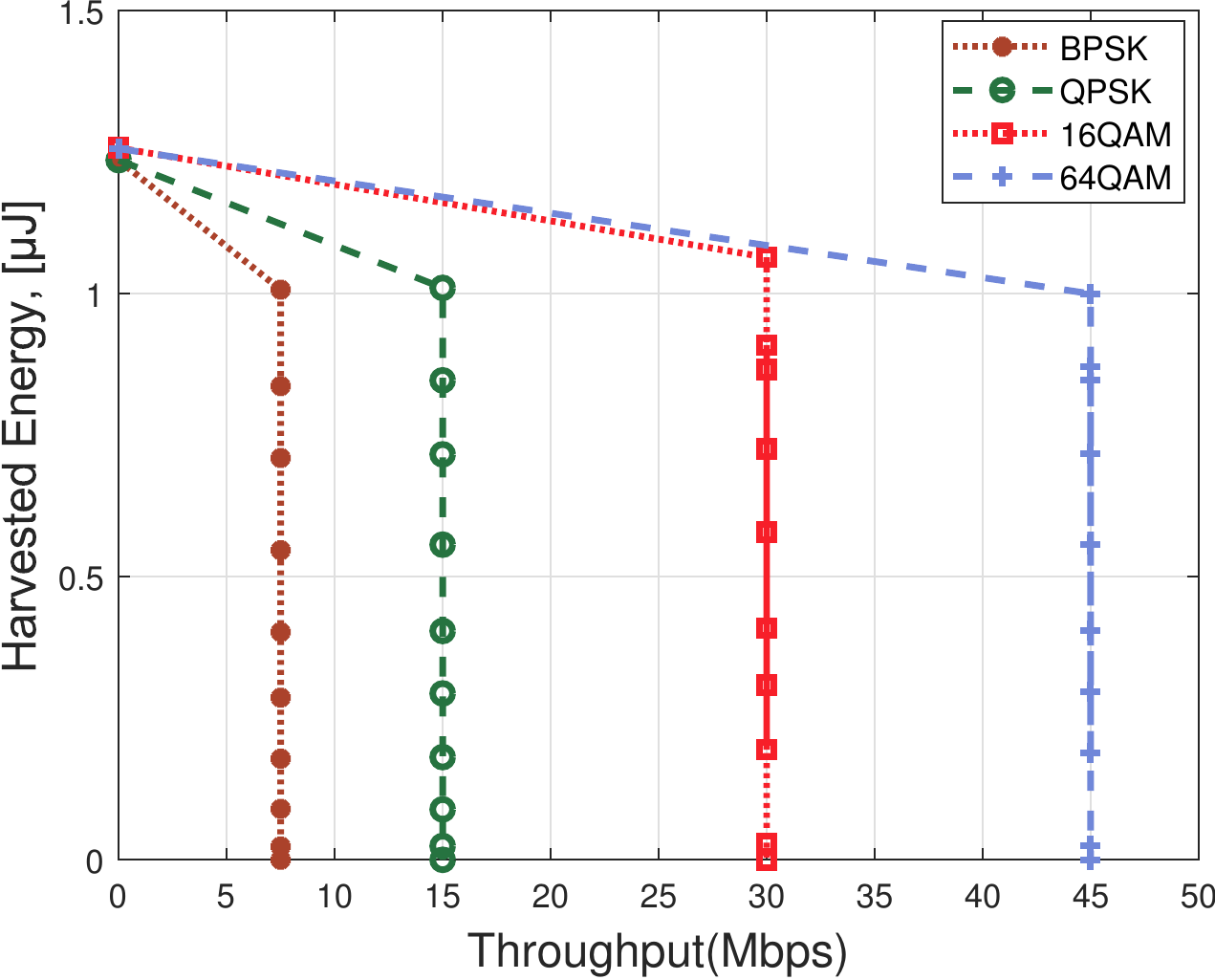}
	\caption{Harvested Energy-Throughput tradeoff at TS and PS receiver architecture.}
	\label{modulation}
\end{figure}

As mentioned in section \ref{system}, the OFDM signal for WIT in our prototype system is designed based on a conventional WiFi 802.11g standard. WiFi signals support several modulation schemes from BPSK to 64 QAM. By changing the modulation scheme of the WIT signal to BPSK, QPSK, 16QAM, and 64QAM,  we evaluated how different modulation schemes affect the performance of the SWIPT system. The test method is the same as the prior subsection and Fig. \ref{modulation} shows the experimental results. 

\par
The results show that changing the modulation does not notably affect the energy harvesting performance but significantly influences the throughput. As mentioned in the prior subsection, the SNR is high enough that the bit-error rate approaches zero even if a higher modulation scheme is used, so the higher the modulation scheme, the larger the energy-throughput region. This observation also allows us to see how the transmission signal design affects the E-T performance of the SWIPT system.

\section{Conclusion and Future Works}\label{conclusion}
This paper described the implementation of a realistic SWIPT prototype and experimentally evaluated how the SWIPT waveform design affects the energy-throughput performance. The experimental results show that a SWIPT signal design that exploits the rectifier nonlinearity has a significant effect on improving the WPT performance of the system and can improve the WIT performance by using various modulation schemes. We have also confirmed that systematic transmission signal design methods such as \textit{superposition} and \textit{time-sharing} methods can significantly improve the energy-throughput (E-T) performance of the realistic SWIPT system. We have also shown that SWIPT system constituents such as the optimal signal design techniques, receiver architectures, and the ID receiver performance are closely related and affect the E-T performance of the overall system. Therefore, the design of the SWIPT system requires a comprehensive consideration of those three constituents. 
\par
Many interesting future works can arise from the observation of this paper. We could expect additional throughput and further expanding the E-T region by using higher modulation because a sufficiently large SNR is ensured in the existing prototype. We could also consider different types of ID receivers which have low receiver sensitivity and examine the optimal signal design and receiver architecture accordingly. Overall, these observations provide useful insights for future practical SWIPT systems and signal design.

\bibliographystyle{IEEEtran}
\bibliography{jhlib} 

%

\end{document}